\documentclass{anabs}
\usepackage{graphicx}
\usepackage{times}
\usepackage{amssymb}
\Volume{1-2}              
\Year{2003}              
\Month{02}               
\Pagespan{000}{000}      

\begin{document}
\lhead[\thepage]{P. Ranalli: Faintest star forming galaxies}
\rhead[Astron. Nachr./AN~{\bf 324} (2003) 1/2]{\thepage}
\headnote{Astron. Nachr./AN {\bf 324} (2003) 1/2, 000--000}

\title{The Faintest Star Forming Galaxies}

\author{Piero Ranalli
}
\institute{
  Dipartimento di Astronomia, Universit\`a di Bologna,
  via Ranzani 1, I--40127 Bologna, Italy
}

\correspondence{ranalli@bo.astro.it}

\maketitle

\newcommand{\Hii}{\ion{H}{ii} }
\newcommand{\Log}{\mbox{Log}}
\newcommand{\XMM}{XMM-{\em Newton} }
\newcommand{\chandra}{{\em Chandra} }
\newcommand{\flux}{{\sc Flux} }
\newcommand{\lum}{{\sc Lum.} }
\newcommand{\lognlogs}{Log~$N$--Log~$S$ }
\newcommand{\lognlogsa}{Log~$N$--Log~$S$}

\section{X-rays from the radio sub-mJy galaxies: two megaseconds of \chandra in the HDFN}

We searched for X-ray detections of star forming galaxies at high
redshift in the 2 Ms \chandra observation of the Hubble Deep Field
North (HDFN).  Star forming galaxies with $0.2\lesssim z\lesssim 1.3$
were selected from the deep radio surveys in the HDFN (Richards et
al.~1998, AJ 116, 1039; Richards 2000, ApJ 533, 611; Garrett 2000,
 A\&A 361, L41); our selection criterium has
been to include all galaxies with Spiral or Irregular morphologies,
known redshifts and no AGN signatures in their optical spectra. 
From a preliminary data reduction, ten sources 
were detected in
the X-rays; for them we determined rest-frame 0.5-2.0 and 2.0-10 keV
fluxes and best-fit X-ray slopes with the same method described in
Ranalli et al.~(2002, A\&A in press).

X-ray and radio fluxes and luminosities (Fig.~1) of these 10
high redshift objects follow the same linear relation which
holds for nearby galaxies and allows the use of the X-ray
luminosity as a Star Formation Rate indicator
(Ranalli et al.~2002, Proc.\ Symp.\ ``New Visions of
the X-ray Universe'', ESTEC 2001, {\em astro-ph/0202241}).
  With fluxes of
the order of a few $10^{-17}$ erg s$^{-1}$ cm$^{-2}$, these are the
faintest normal galaxies ever detected in the X-rays.

\section{X-ray number counts and background}

We consider the number counts for the radio sub-mJy population associated with
\begin{figure}[!h]    
  \begin{center} 
      \includegraphics[width=0.4\textwidth,height=0.34\textwidth]{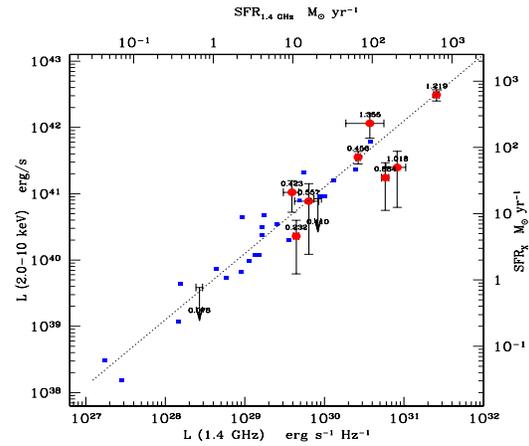}
  \end{center}
  \caption { \small The radio/X-ray luminosity relation for galaxies in the Hubble 
   Deep Field (circles; redshifts are superimposed). 
   Squares: nearby galaxies ($z<0.01$) from Ranalli et al.~(2002).
    Dotted lines:
    best fit relations for nearby galaxies. The same relation also holds for
    fluxes.
  \label{fig_deeplumin}
  }
\end{figure}
faint blue galaxies at high redshift ($22\lesssim V\lesssim 27$, $0.5\lesssim
z\lesssim 1.5$) representing an early era of star formation in the universe.
The deepest radio surveys (at 5 GHz: Fomalont et al.~1991, AJ 102, 1258; at 1.4 GHz:
Richards 2000) give the \lognlogs for this population.

Under the assumption that all the objects are placed at the same redshift (so that
K-corrections are the same), the radio \lognlogs can be converted to X-ray counts
via the radio/X-ray relation.
 We find that the number counts of star forming galaxies should overcome AGN counts 
at fluxes of the order $10^{-17}$ erg s$^{-1}$ cm$^{-2}$. Our prediction for the soft
X-ray counts is fully consistent with the constraints from fluctuation analysis
in the deepest \chandra fields (Miyaji \& Griffiths 2002, ApJ 564, L5). 


\begin{figure}[!h]    
  \begin{center}
  \includegraphics[width=0.4\textwidth,height=0.34\textwidth]{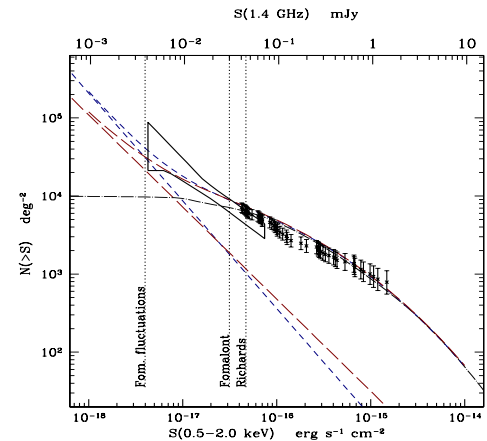}
  \end{center} 
  \caption { \small X-ray counts derived from deep radio \lognlogsa.
    The blue short-dashed and red long-dashed straight lines
  represent X-ray counts derived from the 1.4 GHz (Richards 2000) and 5 GHz 
  (Fomalont et al.~1991) \lognlogs 
  respectively.  Dots: observed X-ray number counts in the 1 Ms \chandra HDFN
  survey (Brandt et al.~2001, AJ 122, 2810). Horn-shaped symbols: results from X-ray fluctuation
  analysis (Miyaji \& Griffiths 2002). Dot-dashed line: 
  number counts from AGN synthesis models
  (Comastri et al.~1995, A\&A 296, 1).  Vertical dotted lines: limiting
  sensitivities for the radio surveys. The sum of galaxies and AGN counts is also shown.}
\end{figure}


\end{document}